\documentclass[article,preprint,groupedaddress,amsmath,amssymbm]{revtex4}
\usepackage{epsfig}
\usepackage{graphicx}

\begin{document}
\title{A quantum bound on the thermodynamic description of gravity}
\author{Shahar Hod}
\address{The Ruppin Academic Center, Emeq Hefer 40250, Israel}
\address{ }
\address{The Hadassah Institute, Jerusalem 91010, Israel}
\date{\today}

\begin{abstract}

\ \ \ The seminal works of Bekenstein and Hawking have revealed that
black holes have a well-defined thermodynamic description. In
particular, it is often stated in the physical literature that black
holes, like mundane physical systems, obey the first law of
thermodynamics: $\Delta S=\Delta E/T_{\text{BH}}$, where
$T_{\text{BH}}$ is the Bekenstein-Hawking temperature of the black
hole. In the present work we test the regime of validity of the
thermodynamic description of gravity. In particular, we provide
compelling evidence that, due to quantum effects, the first law of
thermodynamics breaks
down in the low-temperature regime $T_{\text{BH}}\times
r_{\text{H}}\lesssim ({{\hbar}/{r_{\text{H}}}})^2$ of near-extremal
black holes (here $r_{\text{H}}$ is the radius of the black-hole
horizon).
\newline
\newline
\end{abstract}
\bigskip
\maketitle


\section{Introduction}

It is well known \cite{Bek1} that the thermodynamic description of
mundane physical systems breaks down in the low-temperature regime
$T\sim \hbar/R$ \cite{Noter,Noteunit}, when the characteristic
thermal wavelengths $\lambda_{\text{thermal}}\sim \hbar/T$ are no
longer small on the scale $R$ set by the spatial size of the system.
The physical properties of these low-temperature systems are
dominated by quantum (rather than thermodynamic) effects. Thus, the
thermodynamic description of mundane physical systems is known to be
restricted to the high-temperature regime \cite{Bek1}
\begin{equation}\label{Eq1}
T\times R\gg \hbar\  .
\end{equation}

Interestingly, black holes are unique in this respect. It is well
known that the Bekenstein-Hawking temperature \cite{Bek2,Haw} of the
Schwarzschild black hole is given by $T_{\text{BH}}=\hbar/4\pi
r_{\text{H}}$, where $r_{\text{H}}=2M$ is the radius of the
black-hole horizon. Thus, Schwarzschild black holes are
characterized by the relation $T_{\text{BH}}\times
r_{\text{H}}\sim\hbar$.

Moreover, the Bekenstein-Hawking temperature of Kerr black holes is
given by
\begin{equation}\label{Eq2}
T_{\text{BH}}={{\hbar(r_+-r_-)}\over{4\pi r^2_+}}\  ,
\end{equation}
where $r_{\pm}=M+(M^2-a^2)^{1/2}$ are the black-hole (outer and
inner) horizon radii \cite{Notema}. Thus, rapidly-rotating Kerr
black holes in the near-extremal $r_+-r_-\ll r_+$ regime are
characterized by the strong inequality
\begin{equation}\label{Eq3}
T_{\text{BH}}\times r_+\ll \hbar\  .
\end{equation}
It is quite remarkable that black holes have a well defined
thermodynamic behavior in the low-temperature regime (\ref{Eq3}),
where mundane physical systems are governed by quantum effects and
no longer have a self-consistent thermodynamic description.

One naturally wonders whether the thermodynamic description of black
holes is valid all the way down to the zero temperature $T\times
r_+\to 0$ limit? In order to address this interesting question, we
shall analyze in this paper the regime of validity of the first law
of thermodynamics for black holes.

\section{Black holes and the first law of thermodynamics}

For a closed physical system with a well defined temperature $T$, a
change $\Delta E$ in the energy of the system results with a change
\cite{Lan,Notevol}
\begin{equation}\label{Eq4}
\Delta S={{\Delta E}\over{T}}\
\end{equation}
in the entropy of the system. This famous differential relation,
known as the first law of thermodynamics, is one of the most
important features of the thermodynamic description of mundane
physical systems.

It is often stated in the physical literature (see e.g.,
\cite{Mech}) that black holes, like ordinary thermodynamic systems,
also obey this law. However, the regime of validity of the standard
thermodynamic description of black holes has never been discussed in
the literature. In the present paper we would like to raise the
following intriguing question: Do black holes {\it always} obey the
first law of thermodynamics?

In order to address this interesting question, it proves useful to
examine carefully the assumptions made in the physical literature in
deriving the first law of black-hole thermodynamics. The
characteristic Bekenstein-Hawking entropy of a black hole is given
by a quarter of its horizon area \cite{Bek2,Haw}:
\begin{equation}\label{Eq5}
S_{\text{BH}}={{A}\over{4\hbar}}\  .
\end{equation}
Remembering that the surface area of Kerr black holes \cite{Notekr}
is given by $A=4\pi(r^2_++a^2)=8\pi Mr_+$, one finds \cite{Noteja}
\begin{equation}\label{Eq6}
S_{\text{BH}}={{2\pi M\{M+[M^2-(J/M)^2]^{1/2}\}}\over{\hbar}}\  .
\end{equation}

Consider now a physical process which changes the energy (mass) of
the black hole by a small amount $\Delta E\ll M$ \cite{Noteby}. From
Eq. (\ref{Eq6}) one finds that the resulting change in the
Bekenstein-Hawking entropy of the black hole is given by the rather
cumbersome expression
\begin{equation}\label{Eq7}
{{\hbar}\over{2\pi}}\Delta S_{\text{BH}}=(M+\Delta E) \Big\{M+\Delta
E+\Big[(M+\Delta E)^2-\Big({{J}\over{M+\Delta
E}}\Big)^2\Big]^{1/2}\Big\}-M\Big\{M+\Big[M^2-\Big({{J}\over{M}}\Big)^2\Big]^{1/2}\Big\}\
.
\end{equation}
A careful inspection of Eq. (\ref{Eq7}) reveals that {\it if} $M
\Delta E\ll (r_+-r_-)^2$ then, to leading-order in the small ratio
$M \Delta E/(r_+-r_-)^2$, the changes in the black-hole entropy and
energy are related to each other by the standard first law of
thermodynamics,
\begin{equation}\label{Eq8}
\Delta S_{\text{BH}}={{\Delta E}\over{T_{\text{BH}}}}\  ,
\end{equation}
where $T_{\text{BH}}$ as given by (\ref{Eq2}) is the familiar
Bekenstein-Hawking temperature of the black hole.

On the other hand, in the opposite regime $M \Delta E\gg
(r_+-r_-)^2$ one finds from (\ref{Eq7}) the non-standard relation
\begin{equation}\label{Eq9}
\Delta S_{\text{BH}}={{{\sqrt 8}\pi}\over{\hbar}}M^{3/2}\sqrt{\Delta
E}
\end{equation}
between the changes in the physical parameters (entropy and energy)
of the black hole.

One therefore concludes that, for black holes in Einstein gravity,
the validity of the standard first law of thermodynamics, Eq.
(\ref{Eq8}), is restricted to the regime
\begin{equation}\label{Eq10}
{{(r_+-r_-)^2}\over{r_+}}\gg \Delta E\geq\Delta E_{\text{min}}\  ,
\end{equation}
where $\Delta E_{\text{min}}$ is the smallest possible change in the
energy (mass) of the black hole. Thus, in order to determine the
regime of validity of the law in the context of black-hole physics,
one should first determine the value of the fundamental physical
parameter $\Delta E_{\text{min}}$.

How small can $\Delta E_{\text{min}}$, the minimal change in the
energy (mass) of a black hole, be made? The answer to this question
at the {\it classical} level was given by Christodoulou and Ruffini
\cite{Chr}: the capture of a point particle by a black hole is
characterized by the relation $\Delta E_{\text{min}}=0$ if the
particle is captured at the black-hole horizon from a radial turning
point of its motion. In this scenario the energy (as measured by
asymptotic observers) of the absorbed particle is completely
red-shifted. Substituting $\Delta E_{\text{min}}=0$ into
(\ref{Eq10}) one deduces that, at the classical level, the first law
of thermodynamics may be valid all the way down to the extremal
limit $T_{\text{BH}}\to 0$.

However, as emphasized by Bekenstein in his seminal work
\cite{Bek2}, the classical limit of a perfectly localized particle
(a point particle) is physically unacceptable in a self-consistent
{\it quantum} theory of relativity. In particular, due to the
quantum uncertainty principle \cite{Bor}, the particle cannot be
localized at the black-hole horizon without having a non-zero radial
momentum (kinetic energy). Specifically, the Heisenberg uncertainty
principle sets a lower bound on the smallest possible energy
delivered to the black hole by the captured particle \cite{Bek2}:
\begin{equation}\label{Eq11}
\Delta E_{\text{min}}=2\pi T_{\text{BH}}\  .
\end{equation}
Substituting (\ref{Eq11}) into (\ref{Eq10}) one finds that, within
the framework of a self-consistent quantum theory of gravity, the
validity of the first law of thermodynamics, Eq. (\ref{Eq8}), is
restricted to the regime \cite{Notecaf}
\begin{equation}\label{Eq12}
T_{\text{BH}}\times r_{\text{+}}\gg
\Big({{\hbar}\over{r_{\text{+}}}}\Big)^2\  .
\end{equation}

\section{Summary}

The seminal works of Bekenstein and Hawking have revealed that
gravity has a well-defined thermodynamic description. In particular,
it is often stated in the physical literature that black holes, like
mundane physical systems, obey the first law of thermodynamics:
$\Delta E=T_{\text{BH}}\Delta S$, where $T_{\text{BH}}$ is the
Bekenstein-Hawking temperature of the black hole. In the present
paper we have explored the regime of validity of this law. In
particular, we have shown that,
due to quantum effects, the first law of thermodynamics breaks down
in the low-temperature regime
\begin{equation}\label{Eq13}
T_{\text{BH}}\times r_{\text{H}}\lesssim
\Big({{\hbar}\over{r_{\text{H}}}}\Big)^2
\end{equation}
of near-extremal black holes.

\bigskip
\noindent {\bf ACKNOWLEDGMENTS}
\bigskip

This research is supported by the Carmel Science Foundation. I thank
Yael Oren, Arbel M. Ongo, Ayelet B. Lata, and Alona B. Tea for
stimulating discussions.


\newpage

\begin{thebibliography}{99}

\bibitem{Bek1} J. D. Bekenstein, Phys. Rev. D {\bf 23}, 287 (1981).

\bibitem{Noter} Here $R$ is the effective radius of the
system.

\bibitem{Noteunit} We shall use gravitational units in which $G=c=k_{\text{B}}=1$.

\bibitem{Bek2} J. D. Bekenstein, Phys. Rev. D {\bf 7}, 2333 (1973).

\bibitem{Haw} S. W. Hawking, Commun. Math. Phys. {\bf 43}, 199 (1975).

\bibitem{Notema} Here $M$ and $a\equiv J/M$ are the mass and angular-momentum
per unit mass of the black hole, respectively.

\bibitem{Notevol} Here we refer to closed physical systems with fixed volumes.

\bibitem{Lan} L. D. Landau and E. M. Lifshitz, {\it Statistical
Physics} (Addison-Wesley, Reading, Mass., 1969).

\bibitem{Mech} B. Carter, in {\it Black Holes}, edited by C. M. DeWitt and B. S. DeWitt (Gordon and Breach,
New York, 1973).

\bibitem{Notekr} For concreteness, we shall henceforth consider the physics of
astrophysically realistic Kerr black holes. However, it is worth
emphasizing that our physical conclusions (to be presented below)
can easily be generalized to other types of black holes.

\bibitem{Noteja} Here we have used the relation $J=Ma$ for the
black-hole angular momentum.

\bibitem{Noteby} For example, one can throw into the
black hole a small body whose total energy (as measured by
asymptotic observers) is $\Delta E$.

\bibitem{Chr} D. Christodoulou and R. Ruffini, Phys. Rev. D {\bf 4}, 3552 (1971).

\bibitem{Bor} M. Born, {\it Atomic Physics} (Blackie, London, 1969).

\bibitem{Notecaf} Here we have used the relation
$(r_+-r_-)^2/r_+\sim (T_{\text{BH}}/\hbar)^2r^3_+$ in Eq.
(\ref{Eq10}).

\end{thebibliography}
\end{document}